\documentstyle[12pt]{article}
\date{
      CHIBA-EP-72,
      December  1993
      (hep-th/9401012)  }
 \title{
Flavor-Dependence and Higher Orders of \\
Gauge-Independent Solutions in \\
Strong Coupling Gauge Theory
}
\author{
        Kei-Ichi Kondo,
        \thanks{e-mail : kondo@tansei.cc.u-tokyo.ac.jp;
        kondo@cuphd.nd.chiba-u.ac.jp}
        Takuya Iizuka,
        \thanks{e-mail: tiizuka@cuphd.nd.chiba-u.ac.jp}
        Eiji Tanaka
        \thanks{e-mail: etanaka@cuphd.nd.chiba-u.ac.jp}
        and Toru Ebihara
        \thanks{e-mail: ebihara@cuphd.nd.chiba-u.ac.jp}
        \\
Department of Physics, Faculty of Science \\
         and \\
         Graduate School of Science and Technology,%
         \thanks{mailing address}\\
         Chiba University, Chiba 263, Japan}
\begin{document}
\maketitle
\begin{abstract}
The fermion flavor $N_f$ dependence of non-perturbative
solutions in the strong coupling phase of the gauge theory is
reexamined based on the interrelation between the inversion method
and the Schwinger-Dyson equation approach.
Especially we point out that the apparent discrepancy on the value of
the critical coupling in QED will be resolved by taking into account
the higher order corrections which inevitably lead to the
flavor-dependence.
In the quenched QED,  we conclude that the gauge-independent critical
point $\alpha_c=2\pi/3$ obtained by the inversion method to the
lowest order will be reduced to the result $\alpha_c=\pi/3$ of the
Schwinger-Dyson equation in the infinite order limit, but its
convergence is quite slow. This is shown by adding the
chiral-invariant four-fermion interaction.
\end{abstract}
\newpage
\section{Introduction}
To study the dynamical symmetry breaking, we need some
non-perturbative tool. In recent several years,  the
Schwinger-Dyson (SD) equation has played such a role and much
effort is devoted to solving the  SD  equation for the fermion
propagator in gauge theories.
In QED$_4$ with $N_f$
fermion flavors,  QED$_4[N_f]$,  there exists a critical gauge
coupling $\alpha_c$  for
$\alpha$, above which ($\alpha>\alpha_c$) the chiral symmetry is
spontaneously broken, i.e.,
$\langle \bar \psi \psi \rangle\not=0$ in the absence of bare
fermion mass.
Such non-perturbative solutions of the SD equation ware
able to reveal many interesting features of the strong coupling
phase in gauge theories, see e.g., \cite{Yamawaki91,Kondo92p} for
review.
 However this approach has the following drawbacks.
\begin{enumerate}
\item 
systematic improvement: It is rather difficult to incorporate the
higher order effect systematically.
In contrast to the perturbation theory, it is not necessarily
clear which terms should be taken account of to the next order.
Even if this may be possible, it is quite difficult to
actually solve it without drastic approximations. In essence, the
SD equation for the fermion propagator should be solved
simultaneously with those for the photon propagator and the
vertex function and such an attempt has been done in a restricted
case with a bare vertex \cite{KMN92}.
However the {\em ad hoc} approximation for the vertex
function and  the photon propagator inevitably truncates this
series of hierarchy.
\item 
gauge invariance: Any gauge-independent result has not been obtained
so far. Almost all the analyses of the SD equation have been done
under the choice of the Landau gauge.
 However, qualitatively different and pathological results are
obtained, once the other gauges than the Landau gauge are chosen
\cite{NT89}.
This implies that
the adopted  approximation breaks the gauge invariance which
should hold in gauge theories.
 In other gauges than the Landau gauge, we must take into
account the wavefunction renormalization for the fermion, even in
the quenched approximation \cite{KN89}.  This enforces us to
consider the  vertex correction in the light of the Ward-Takahashi
(WT) identity. Indeed, many physicists have tried to recover the
gauge-invariance by modifying the vertex so as to
satisfy the WT identity.
 However the WT identity does not uniquely determine the vertex
function of the SD equation in the non-perturbative sense
\cite{KK88}, in contrast with the perturbation theory.  We must
impose further conditions or requirements; absence of the kinematical
singularity, multiplicative renormalizability, agreement with the
perturbation theory in the weak-coupling limit and so on. In spite
of these trials, however, completely gauge-independent results  have
not yet been obtained in the framework of the SD equation.
See references \cite{Kondo92,CP93,AGM93,ABGPR93} for the most
recent result.
\end{enumerate}
\par
The inversion method \cite{Fukuda88} is a generalization of
the Legendre transformation in the effective-action formalism
\cite{CJT74} and enables us to calculate the order parameter in the
symmetry-breaking non-perturbative phase in a perturbative way.  The
order parameter in the effective action is uniquely determined by
the form of the source term in the action
$S_J$ from which we start, while in the inversion method it can
be chosen more freely providing that it coincides with the
original action when the source term vanishes.
 In the inversion method, we can extract the non-perturbative
solution by perturbative calculations.
This method
allows us to perform the systematic improvement \cite{Ukita91}.
 By using this method, the strong-coupling phase of  QED is
studied in a gauge-invariant manner \cite{UKF90,Kondo93}.
Accordingly the inversion method is very useful and has
remarkable features.
\par
However it has not attracted so much
attention of most physicists who have been studying the
dynamical symmetry breaking.  For one thing, the
relationship between the inversion method and the SD equation
approach was not necessarily made clear so far; for another
thing, the SD equation approach is more direct to obtain the
physical quantities, such as the fermion mass and the pion decay
constant.
\par
Moreover there exists an apparently contradicting result on
the critical coupling already in QED between the inversion method and
the SD equation approach.
\par
 In this letter we
show that there is no contradiction between the result of the
inversion method and the  SD equation approach. Our systematic
treatment of the solution of  the SD equation and its
comparison with the inversion result suggests that the
correct gauge-independent critical coupling in the quenched
limit should be given by $\alpha_c=\pi/3$.

 \par
\section{SD equation and inversion method}
First we recall the result of the SD equation.  The SD equation is
also derived from the inversion method as follows. After introducing
the source term,
\begin{equation}
S_J = \int d^4x \int d^4y J(x,y) \bar{\psi}(x) \psi(y),
\end{equation}
into the  original action of (euclidean) QED$_4[N_f]$:
\begin{eqnarray}
S_{QED} &=& \int d^4x  {\cal L}_{QED},
\nonumber\\
{\cal L}_{QED} &=&
 - {1 \over 4} F_{\mu \nu}F^{\mu \nu}
 + \bar{\psi}^a (i \rlap{$\partial$}/ - m_0) \psi^a
 + e_0 \bar{\psi}^a \gamma_{\mu}\psi^{a}A^{\mu} ,
\ (a = 1,..., N_f),
\end{eqnarray}
the inversion process  with respect
to the general bilocal source
$J(x,y)$ yields, to the lowest order ${\cal O}(e_0^2)$,
the equation:
\begin{eqnarray}
J(p) = - S(p)^{-1}  +  S_0(p)^{-1}
 -   e_0^2 \int  {d^4 q \over  (2\pi)^4}
  \gamma_\mu S(q) \gamma_\nu D_0^{\mu\nu} (q - p) ,
\label{SDeq}
\end{eqnarray}
where $J(p)$ is the Fourier transformation of $J(x-y)$, and $S_0(p)$
and
$D_0^{\mu\nu} (k)$ denote the bare fermion propagator and the bare
photon propagator, respectively.
By putting $J = 0$, the {\em quenched} ladder SD equation for the
fermion propagator
$S(p)$ is obtained.
The quenched ladder SD equation in the Landau gauge
\cite{Miransky85} leads to the critical coupling
\footnote{The value $\alpha_c(\Lambda)$ of the critical coupling
obtained from the  quenchd ladder SD equation for the fermion
propagator $S(p) = [A(p^2) \gamma^\mu p_\mu  - B(p^2)]^{-1}$ depends
on both the gauge parameter
$\xi$ and the ultraviolet cutoff $\Lambda$ \cite{NT89,KK88}.
For $\xi \le -3$ there is no nontrivial solution:
$B(x) \equiv 0$.
For $0>\xi>-3$, $\alpha_c(\Lambda)$ exists, but $\alpha_c(\Lambda)
\downarrow 0$ as $\Lambda \uparrow \infty$.
In the Landau gauge $\xi=0$, $\alpha_c(\Lambda) \rightarrow
 \pi/3$ as $\Lambda \uparrow \infty$.
For  $\xi>0$, $\alpha_c(\Lambda) \rightarrow
\alpha_c^\infty > \pi/3$ as $\Lambda \uparrow \infty$ and
$\alpha_c^\infty$ increases monotonically in $\xi$.
Note that $A(p^2) > 1$ for $\xi>0$ and  $0<A(p^2) < 1$ for $\xi<0$.
The quenched ladder SD equation with the bare vertex is consistent
with the WT identity only in the Landau gauge where $A(p^2) \equiv
1$.}:
\begin{equation}
\alpha_c(0) = {\pi \over 3} =  1.041975 \cdots .
\label{cpNf0}
\end{equation}
\par
On the other hand, the lowest order inversion
method applied to the chiral condensates
$\langle~\bar{\psi}\psi\rangle$ of QED$_4[N_f=1]$ leads to the
gauge-independent critical point
\footnote{This value is obtained if we put the same ultraviolet
cutoff for fermion momentum $\Lambda_f$ and photon one $\Lambda_p$
in $\alpha_c=2\pi/(3\eta)$ with $\eta=\Lambda_p^2/\Lambda_f^2$
\cite{UKF90}.
Regularization independence of the critical coupling will be
discussed in a forthcomming paper in detail.}:
\begin{equation}
\alpha_c = {2 \pi \over 3} = 2.094395...,
\label{cpNf1}
\end{equation}
as shown in \cite{UKF90}.
The critical point $\alpha_c$ depends in general on the number of
fermion flavors $N_f$, $\alpha_c(N_f)$ \cite{KKM89,KN91,OJ90}.
To the lowest order inversion, however, the
critical coupling $\alpha_c$ of QED$_4[N_f]$ is independent of the
flavor number of fermion, $N_f$, as pointed out in \cite{Kondo93}.
\par
 The critical coupling eq.~(\ref{cpNf1}) is twice as large as the
result eq.~(\ref{cpNf0}) of the  quenched ladder SD equation
eq.~(\ref{SDeq}) in the Landau gauge. Then the result
eq.~(\ref{cpNf1}) seems to contradict with the inversion scheme.
 Because
 the result eq.~(\ref{cpNf1}) follows  from a special case, i.e.,
the uniform source $J$ with $J(x,y)=J\delta^4(x-y)$ in the same
inversion procedure in which case we have the relation after
inversion:
\begin{eqnarray}
J
&=& {4\pi^2 \over N_f \Lambda_f^2}
\left[ 1 - {\alpha\over \alpha_c} \right]
\langle \bar \psi \psi \rangle
\nonumber\\
&&+  {64 \pi^6 \over \eta N_f^3 \Lambda_f^8}
\langle \bar \psi \psi \rangle^3
\ln ^2 \left({16\pi^4 \over N_f^2\Lambda_f^4 \Lambda_p^2}
\langle \bar \psi \psi \rangle^2 \right)
\nonumber\\
&&+ {\cal O}(\langle \bar \psi \psi \rangle^3
\ln \langle \bar \psi \psi \rangle^2).
\label{inversion}
\end{eqnarray}
\par
 Rather the inversion result  eq.~(\ref{cpNf1}) is in
good agreement with that of the unquenched ladder SD equation (in
the Landau gauge) for one fermion  flavor $N_f$=1
\cite{KKM89,KN91,KN92,OJ90,Gusynin90}:
\begin{equation}
\alpha_c(1) = 2.00,
\label{cp2}
\end{equation}
in the presence of the 1-loop vacuum polarization for the
photon propagator
\begin{equation}
D_{\mu \nu}(k) = {1 \over k^2[1+\Pi^{(1)}(k^2)]}
  \left( \delta_{\mu \nu}-{k_\mu k_\nu \over k^2} \right),
  \
  \Pi^{(1)}(k^2) = {N_f \alpha \over 3\pi} \ln {\Lambda^2 \over k^2},
\end{equation}
where $\Lambda$ is the ultraviolet cutoff.
\par
 How can we make a compromise among these
results? Or, different formalism may give the different critical
point?
 How can the effect of photon vacuum polarization be included
and  the unquenched result be extrapolated to the quenched
limit, $\Pi^{(1)}(k^2) \rightarrow 0$, i.e.,
$D^{\mu \nu}(k) \rightarrow D_0^{\mu \nu}(k)$?
\par
We clarify these points based on the systematic
treatment of the SD equation \cite{KN91,Kondo91,Kondo91b}.

\par
\section{Analysis}
\par
All the above statements will be made clear if we include
the chiral-invariant four-fermion interaction \cite{NJL61} (gauged
Nambu--Jona-Lasinio model \cite{BLL86}) as
\begin{eqnarray}
{\cal L}_{4F}  =
{1 \over 2} G_0 [(\bar{\psi}^a\psi^{a})^{2}
 - (\bar{\psi}^a \gamma_5\psi^{a})^{2}] , \ (a = 1, ..., N_f) ,
\end{eqnarray}
and consider the critical line in the two-dimensional space of
coupling constants $(e_0,G_0)$ instead of the critical point.
We define the dimensionless four-fermion coupling $g$ by
$g \equiv {N_f G_0 \Lambda^2 \over 4\pi^2}$,
so that the critical four-fermion coupling constant $g_c$ is kept
fixed: $g_c=1$ for arbitrary $N_f$ (in the chain approximation).

\par
In the previous paper \cite{Kondo93} the inversion method was
applied to the gauged Nambu--Jona-Lasinio (NJL) model
which reduces to QED by switching off the chiral-invariant
four-fermion interaction.
\footnote{Within the framework of the SD equation, an approach
to recover the gauge invariance approximately was tried for the
gauged NJL model in \cite{Appelquistetal91}.}
To the lowest order, we have obtained the
gauge-independent results:
\begin{enumerate}
\item 
 the critical line separating the spontaneous chiral-symmetry
breaking phase from the chiral symmetric one,
\item 
 the mean-field value $1/2$ for the critical
exponent of the chiral order parameter $\langle~\bar{\psi}\psi\rangle$,
\item 
 the large anomalous dimension $\gamma_m=2$ for the
composite operator $\bar{\psi}\psi$ on the whole critical line.
\end{enumerate}
 These results are in good agreement with the previous results of the
SD equation which includes the 1-loop vacuum polarization to the
photon propagator in QED \cite{Kondo91,Kondo91b}. However
the critical line obtained from the SD equation deviates from the
inversion result in the region of the strong gauge coupling,
$\alpha = e_0^2/4\pi > 4$ where the bare coupling of the
four-fermion interaction is negative \cite{Kondo91b}.  This indicates
necessity of incorporating the higher orders in the inversion
calculation.
\par
{}From the SD equation, the critical line of the gauged NJL
model is shown to take the  following form \cite{Kondo91,Kondo91b}.
\begin{eqnarray}
g  = {1+\sum_{n=1}^{\infty} a_n/z_0^n \over
 1+\sum_{n=1}^{\infty} b_n/z_0^n},
\end{eqnarray}
where
\begin{eqnarray}
a_1 &=& -\sigma^2-1,
 \nonumber\\
a_2 &=& {1 \over 2} \sigma^2(\sigma^2-2\sigma+2),
 \nonumber\\
b_1 &=& -(\sigma-1)^2,
 \nonumber\\
a_{n( \ge 3)} &=& {(-1)^n \over n!} \sigma^2[P_n(\sigma)+n]
 \prod_{i=2}^{n-1} P_i(\sigma),
 \nonumber\\
b_{n( \ge 2)} &=& {(-1)^n \over n!} (\sigma-n)^2
 \prod_{i=2}^{n} P_i(\sigma),
\end{eqnarray}
and
\begin{eqnarray}
P_i(\sigma) := \sigma^2-2(i-1)\sigma+(i-1)(i-2),
\end{eqnarray}
with
\begin{eqnarray}
\sigma = {9 \over 4N_f},
 \ z_0 :=  {3\pi \over N_f \alpha}.
\end{eqnarray}
This is derived from the asymptotic solution of the SD equation
\cite{KN91} with an expansion parameter $z_0^{-1}$.
Up to the order ${\cal O}(z_0^{-10})$, the critical line is
written as
\begin{eqnarray}
g &=& 1 - 2 \lambda
  - \left(1-{2 \over \sigma}\right)
  \lambda^2
  - 2\left(1-{3 \over \sigma}+{2 \over \sigma^2}\right)
  \lambda^3
\nonumber\\
&&  - \left(5-{22 \over \sigma}+{30 \over \sigma^2}-{12 \over \sigma^3}
\right)  \lambda^4
  - 2\left(1-{2 \over \sigma}\right)^2
  \left(7-{14 \over \sigma}+{6 \over \sigma^2}\right)
  \lambda^5
\nonumber\\ &&
- 2\left(1-{2 \over \sigma}\right)^2
 \left(21-{79 \over \sigma}+{90 \over \sigma^2}-{30 \over \sigma^3}\right)
  \lambda^6
\nonumber\\ &&
- 4\left(1-{2 \over \sigma}\right)^2
 \left(33-{187 \over \sigma}+{379 \over \sigma^2}
 -{318 \over \sigma^3}+{90 \over \sigma^4}\right)
  \lambda^7
\nonumber\\ &&
- \left(1-{2 \over \sigma}\right)^2
 \left(429-{3304 \over \sigma}+{9960 \over \sigma^2}
 -{14508 \over \sigma^3}+{9996 \over \sigma^4}
 -{2520 \over \sigma^5}\right)
  \lambda^8
\nonumber\\ &&
- 2 \left(1-{2 \over \sigma}\right)^2
 \Biggr( 715-{7048 \over \sigma}+{28760 \over \sigma^2}
 -{61756 \over \sigma^3}+{72708 \over \sigma^4}
\nonumber\\ &&
\
-{43608 \over \sigma^5}+{10080 \over \sigma^6}\Biggr)   \lambda^9
\nonumber\\ &&
- 2 \left(1-{2 \over \sigma}\right)^2
 \Biggr( 2431-{29479 \over \sigma}+{153724 \over \sigma^2}
 -{445060 \over \sigma^3}+{766920 \over \sigma^4}
\nonumber\\ &&
\
-{776940 \over \sigma^5}+{420120 \over \sigma^6}-{90720 \over \sigma^7}
 \Biggr)
  \lambda^{10}
\nonumber\\ &&
 + {\cal O}\left( \lambda^{11} \right),
\label{criticallineex}
\end{eqnarray}
where
\begin{equation}
\lambda := {\sigma \over z_0} = {3\alpha \over 4\pi}.
\end{equation}
Now this expression allows us to take the quenched limit $N_f
\rightarrow 0$:
\begin{eqnarray}
g &=& 1 - 2 \lambda - \lambda^2 - 2\lambda^3 -5 \lambda^4 -
14\lambda^5
 \nonumber\\ &&
 - 42 \lambda^6
 - 132 \lambda^7 - 429 \lambda^8  - 1430 \lambda^9  - 4862 \lambda^{10}
 - {\cal O}(\lambda^{11}),
\label{clqex}
\end{eqnarray}
with $\lambda$ being kept fixed, although both $\sigma$ and $z_0$
have singular $N_f$-dependence. Therefore we are lead to conclude
that the critical line
eq.~(\ref{criticallineex})
has the correct quenched limit
\footnote{Similar argument can be done for the
QCD-like gauged NJL model, as already pointed out in
\cite{KSY91,Kondo92p}.}
\cite{KMY89}:
\begin{equation}
g = {1 \over 4}\left(1+\sqrt{1-\lambda/\lambda_c}\right)^2,
 \lambda_c = {3\alpha_c \over 4\pi} = {1 \over 4},
\label{clq}
\end{equation}
since eq.~(\ref{clq}) is expanded as
\begin{equation}
g = 1-2\lambda - \sum_{n=2}^\infty {\prod_{k=1}^{n-1}(4k-2) \over
 n!} \lambda^n.
\end{equation}
To the lowest order ${\cal O}(\alpha)$, the critical line
eq.~(\ref{criticallineex}) is given   by
\begin{equation}
 g = 1 - {3\alpha \over 2\pi}.
\label{line1}
\end{equation}
This is independent of the fermion flavor and completely
agrees with the lowest order inversion result \cite{Kondo93}.
This implies the critical coupling $\alpha_c=2\pi/3$ in pure QED
($g=0$) which is nothing but the result of
Ukita, Komachiya and Fukuda\cite{UKF90}.
\par
The complete agreement of the SD result  eq.~(\ref{line1}) with the
inversion result in the lowest order ${\cal O}(\alpha)$ is somewhat
surprising.  Because in the analytical treatment of the SD
equation the standard approximation
$
\Pi^{(1)}((p-q)^2)   \cong \Pi^{(1)}({\rm max}\{p^2,q^2\})
$
is adopted in order to perform the angular
integration and to conclude no wavefunction renormalization in
the Landau gauge as in the quenched case, while there is no
approximation in the inversion method to this order.
However the critical coupling is insensitive to this
approximation \cite{KN92}, although numerical
calculation of the SD equation leads to slightly different value
under different approximations \cite{AGJ92}. For $N_f=1$, our
result shows the critical point
 up to the order ${\cal O}(\alpha^{10})$  (see Table 1)
\begin{equation}
 \alpha_c(1) = {\pi \over 3} \times 1.90942 = 1.9995.
\end{equation}
\par
In order to obtain $N_f$-dependent result, we must take into
account at least the next order ${\cal O}(\alpha^2)$.
Actually, the critical line to the order ${\cal O}(\alpha^2)$ is
given by
\begin{equation}
 g = 1 - {3 \over 2} {\alpha \over \pi}
 - {9 \over 16} \left(1-{8 \over 9} N_f \right)
 \left( {\alpha \over \pi} \right)^2.
\label{criticalline2}
\end{equation}
{}From the viewpoint of the inversion, the different
$N_f$-dependence does appear in the next order from a vacuum
diagram with two fermion loops, see Figure 1 of \cite{UKF90}.
It is worth remarking that the
numerical result of the SD equation, $\alpha_c=2.00$ for $N_f=1$
\cite{KKM89}, is obtained from the above equation by solving the
second order algebraic equation:
$
0 = 1 - 2\lambda -(1-{8 \over 9}N_f) \lambda^2,
$
which has indeed a positive solution at $\alpha_c=2.03924$ for
$N_f=1$. The ratio of the second order term to the first order
one is $(1-{8 \over 9}N_f)/4=1/36=0.0277$ for $N_f=1$ at
$\alpha=2\pi/3$. This is why the lowest order result eq.~(\ref{cpNf1}) is
very close to the value eq.~(\ref{cp2}).  Therefore this coincidence
for $N_f=1$ is an accidental one due to the negligible second
order term.
However, the coefficient of the ${\cal O}(\alpha^2)$ term in
eq.~(\ref{criticalline2}) changes the sign for $N_f > 9/8$.  Hence $g
\nearrow +\infty$ as $\alpha \rightarrow +\infty$, which contradicts
with the SD equation result:
$g \searrow -\infty$ as $\alpha \rightarrow +\infty$
\cite{Kondo91b,Rakow91}.
Nevertheless the critical line eq.~(\ref{criticalline2}) can give
good estimate at least for not so large $\alpha$:
$\alpha \le {\cal O}(\alpha_c)$.
Actually the result in the case of $N_f=2$ should be compared with
the value $\alpha_c(2)=2.5$ obtained from the numerical solution of
the SD equation \cite{KKM89,KN91,KN92}.
\par
We must mention the behavior of the series in more higher orders.
For $N_f>0$, it is known \cite{Kondo91,Kondo92p} that the series
eq.~(\ref{criticallineex})
is not convergent and at most strong asymptotic series (hence Borel
summable) \cite{RS78}, while in the quenched case the series
eq.~(\ref{clqex}) converges to eq.~(\ref{clq}) as
$N \rightarrow \infty$.
Therefore the critical coupling constant $\alpha_c$ approaches to
the optimal value at a certain $N$ and then deviates from it, which
can be seen in Table 1.
However, for large $N_f$ the asymptotic series fails to give
good estimate on the critical coupling, since the expansion
parameter $z_0^{-1}$ should be small: $z_0^{-1}=N_f \alpha/(3\pi)<1$,
as already pointed out \cite{KN91}.
\par
In Table 1, it is demonstrated that all order results
are needed in order to recover the correct
critical coupling
$\alpha_c(0)=\pi/3$ of the quenched limit from the unquenched SD
equation, since the convergence to $\alpha_c(0)=\pi/3$ is
very slow;
$\alpha_c(0)=(\pi/3)\times 1.00957$ to the order
${\cal O}(\alpha^{500})$.

\par
\section{Conclusion}
For QED$_4[N_f]$, it has been shown that the
flavor $N_f$-independent critical value
$\alpha_c=2\pi/3$ is a common value to the "lowest order"
\footnote{The "order" of the result obtained from
the SD equation is defined in correspondence with that of the
inversion method, see eq.~(\ref{criticallineex}).}
${\cal O}(\alpha)$ in the SD equation approach as well as the
inversion method.  The key point lies in the fact that the
$N_f$-dependence appears in the next order ${\cal O}(\alpha^2)$ for
the first time. Therefore it is merely an accidental coincidence that
the lowest order inversion result $\alpha_c=2\pi/3$ is nearly equal
to the value $\alpha_c=2.00$ obtained from the unquenched ladder SD
equation for
$N_f=1$.
\par
The quenched case, $N_f=0$, can be extrapolated from the expression
eq.~(\ref{criticallineex}) which we have obtained from the SD
equation. However, the convergence to $\alpha_c(0)=\pi/3$ is
very slow as demonstrated in Table 1.
Therefore, in order to obtain the quenched
result eq.~(\ref{cpNf0}) in the inversion method, we need to include
all the orders of vacuum diagrams with only one fermion loop
(with no further internal fermion loop). This subtlety seems to
reflect the essential singularity in the Miransky scaling
\cite{Miransky85} in the quenched QED. To obtain such a correct
{\it gauge-independent} quenched result, we have to sum up all the
order results of the inversion.
\par
To really confirm our claims, anyway it is necessary to perform
the calculation to higher orders in the inversion. It is easily
shown that the gauge-independence is guaranteed also
in the higher orders. Then it is important to study the
next order ${\cal O}(\alpha^2)$ of the inversion method in order to
determine whether or not the gauge-independent inversion result
really agrees with the SD equation result in the Landau gauge
eq.~(\ref{criticalline2}).
Details on this point will be reported in a subsequent paper.
\par
\vskip 1cm
\leftline{\bf Acknowledgments}
The first author (K.-I. K.) would like to thank David Atkinson,
Pieter Maris and Manuel Reenders for interesting discussions and
all members of the Institute for Theoretical Physics in
University of Groningen for their warm hospitality where a part
of this work was performed.
He wishes to thank Netherlands Organization for
Scientific Research NWO (Nederlandse organisatie voor
Wetenschappelijk Onderzoek) and the Japan Society for the
Promotion of Sciences JSPS for financial supports which enabled
him to stay in Groningen for two months (September and October
1993).

\newpage
\begin{table}
\begin{center}
\begin{tabular}{rccc}                               \hline
 order $N$  &   $N_f=0$  &   $N_f=1$   &   $N_f=2$   \\ \hline
  $1$   &    $2.0$   &    $2.0$    &    $2.0$    \\
  $2$   & $1.65685$  & $1.94733$   & $2.71849$   \\
  $3$   & $1.50434$  & $1.92136$   & $3.08042$   \\
  $4$   & $1.41548$  & $1.91234$   & $2.46357$   \\
  $5$   & $1.35635$  & $1.91018$   & $2.87205$   \\
  $6$   & $1.31374$  & $1.90964$   & $2.52280$   \\
  $7$   & $1.28136$  & $1.90947$   & ---         \\
  $8$   & $1.25579$  & $1.90943$   & $2.33712$   \\
  $9$   & $1.23501$  & $1.90942$   & ---         \\
 $10$   & $1.21775$  & $1.90942$   & $2.00813$   \\
 $15$   & $1.1616~$  & $1.90938$   & ---         \\
 $20$   & $1.13025$  & $1.91519$   & $0.92838$   \\
 $25$   & $1.10993$  & $1.69725$   & ---         \\
 $30$   & $1.09556$  & ---         & $0.56558$   \\
 $35$   & $1.08481$  & $1.12770$   & ---         \\
 $40$   & $1.07643$  & $*$         & $*$         \\
 $50$   & $1.06413$  & $*$         & $*$         \\
 $60$   & $1.05549$  & $*$         & $*$         \\
 $70$   & $1.04906$  & $*$         & $*$         \\
 $80$   & $1.04406$  & $*$         & $*$         \\
 $90$   & $1.04006$  & $*$         & $*$         \\
$100$   & $1.03678$  & $*$         & $*$         \\
$200$   & $1.02077$  & $*$         & $*$         \\
$300$   & $1.01478$  & $*$         & $*$         \\
$400$   & $1.01158$  & $*$         & $*$         \\
$500$   & $1.00957$  & $*$         & $*$         \\ \hline
\end{tabular}
\end{center}
Table 1. Critical coupling constant $\alpha_c$ in unit of
$\pi/3$ in QED$_4[N_f]$,
 obtained by solving the algebraic equation
eq.~(\ref{criticallineex})
$g=f(\alpha)=0$ up to the order
${\cal O}(\alpha^N)$.
 To the lowest order ${\cal O}(\alpha)$, $\alpha_c=2\pi/3$
irrespective of the fermion flavor $N_f$.
 In the quenched case ($N_f=0$) the critical coupling  approaches the
value, $\alpha_c=\pi/3$ for large $N$.
 The symbol $-$ stands for no real positive solution to the
 algebraic equation $f(\alpha)=0$ and
 $*$ implies that we have no available data.
\end{table}


\newpage
\begin{thebibliography}{99}
%
\bibitem{Yamawaki91}
  K. Yamawaki,
  {\it Top-Mode Standard Model,
  in Proc. 1990 International Workshop on Strong Coupling
  Gauge Theories and Beyond, Nagoya, July 28-31, 1990}, eds. T.
Muta
  and K. Yamawaki (World Scientific Pub. Co., Singapore, 1991).
%
\bibitem{Kondo92p}
  K.-I. Kondo,
  {\it Triviality Problem and Schwinger-Dyson Equation Approach,
  in Dynamical Symmetry Breaking},  ed. K. Yamawaki (World
Scientific Pub. Co.,  Singapore, 1992).
%
\bibitem{KMN92}
  K.-I. Kondo, H. Mino and H. Nakatani,
  Mod. Phys. Lett. {\bf A7} (1992) 1509.
%
\bibitem{NT89}
  R.W. Haymaker, Acta Physica Polonica {\bf B13} (1982) 575;
  D. Atkinson and P.W. Johnson, J. Math. Phys. {\bf 28} (1987)
2488;
  T. Nonoyama and M. Tanabashi, Prog. Theor. Phys. {\bf 81}
(1989) 209;
  K-I. Aoki, M. Bando, K. Hasebe, T. Kugo and H. Nakatani,
  Prog. Theor. Phys. {\bf 82} (1989) 1151.
%
\bibitem{KN89}
  K.-I. Kondo and H. Nakatani,
  Mod. Phys. Lett. {\bf A4} (1989) 2155.
%
\bibitem{KK88}
  K.-I. Kondo and Y. Kikukawa,
  Nagoya Univ. Preprint, DPNU-88-20, unpublished;
  H. Mino, in the Proceedings of the 1988 International
Workshop on  New Trends in Strong Coupling Gauge Theories,
Nagoya, Aug. 24-27, eds. by M. Bando, T. Muta and K. Yamawaki
(World Scientific Co., Singapore, 1989).
%
\bibitem{Kondo92}
  K.-I. Kondo, Intern. J. Mod. Phys. {\bf A 7} (1992) 7239.
%
\bibitem{CP93}
  D.C. Curtis and M.R. Pennington,
  Phys. Rev. D 44 (1991) 536;
  Univ. of Durham, preprint, DTP-93/20.
%
\bibitem{AGM93}
  D. Atkinson, V.P. Gusynin and P. Maris,
  Phys. Lett. {\bf B 303} (1993) 157.
%
\bibitem{ABGPR93}
  D. Atkinson, J.C.R. Bloch, V.P. Gusynin, M.R. Pennington and
M. Reenders,
  Univ. of Groningen/Durham, preprint, RUG-TH-930901/DTP-93/62.
%
\bibitem{CJT74}
  J.M. Cornwall, R. Jackiw and E. Tomboulis,
  Phys. Rev. {\bf D10} (1974) 2428.
%
\bibitem{Fukuda88}
  R. Fukuda,
Phys. Rev. Lett. {\bf 61} (1988) 1549;
{\it Novel Use of the Effective Action,
in Dynamical Symmetry Breaking},
ed. K. Yamawaki (World Scientific Pub. Co.,  Singapore, 1992).
%
\bibitem{Ukita91}
  M. Ukita,
  {\it Functional Legendre Transformation and Physical
Properties of Field Theoretical Systems},
  Ph.D. thesis, Keio University, 1991.
%
\bibitem{UKF90}
  M. Ukita, M. Komachiya and R. Fukuda,
Intern. J. Mod. Phys. {\bf A  5} (1990) 1789.
%
\bibitem{Kondo93}
  K.-I. Kondo,
  Chiba Univ. Preprint, CHIBA-EP-69, hep-th/9305186,
  Mod. Phys. Lett. {\bf A 8} (1993) 3031.
%
\bibitem{Miransky85}
  V.A. Miransky,
  Nuovo Cimento {\bf 90A} (1985) 149.
%
\bibitem{KKM89}
   	K.-I. Kondo, Y. Kikukawa and H. Mino,
    Phys. Lett. {\bf B 220} (1989) 270.
%
\bibitem{KN91}
   	K.-I. Kondo and H. Nakatani,
Nucl. Phys. {\bf B 351} (1991) 236.
%
\bibitem{KN92}
   	K.-I. Kondo and H. Nakatani,
Prog. Theor. Phys.  {\bf 88} (1992) 737.
%
\bibitem{OJ90}
   	J. Oliensis and P.W. Johnson,
    Phys. Rev. {\bf D 42} (1990) 656.
%
\bibitem{Gusynin90}
  V.P. Gusynin,
  Mod. Phys. Lett. {\bf A 5} (1990) 133;
  Ukrainian Journal of Physics, {\bf 35} (1990),
No.7.
%
\bibitem{Kondo91}
  K.-I. Kondo, Nucl. Phys. {\bf B 351} (1991) 259.
%
\bibitem{Kondo91b}
  K.-I. Kondo, Intern. J. Mod. Phys. {\bf A 6} (1991) 5447.
%
\bibitem{NJL61}
  Y. Nambu and G. Jona-Lasinio,
  Phys. Rev. {\bf 122} (1961) 345.
%
\bibitem{BLL86}
  W.A. Bardeen, C. Leung and S. Love,
  Phys. Rev. Lett. {\bf 56} (1986) 1230;
  Nucl. Phys. {\bf B 273} (1986) 649.
%
\bibitem{Appelquistetal91}
T. Appelquist, U. Mahanta, D. Nash and L.C.R. Wijewardhana,
Phys. Rev. {\bf D 43} (1991) 646.
%
\bibitem{KMY89}
  K.-I. Kondo, H. Mino and K. Yamawaki,
  Phys. Rev. {\bf D39} (1989) 2430;
  T. Appelquist, M. Soldate, T. Takeuchi and L.C.R. Wijewardhana,
  {\it in Proc. Johns Hopkins Workshop on
  Current Problems in Particle Theory 12, Baltimore, June 8-10, 1988,}
  eds. G. Domokos and S. Kovesi-Domokos
  (World Scientific Pub. Co., Singapore, 1988).
%
\bibitem{KSY91}
  K.-I. Kondo, S. Shuto and K. Yamawaki,
  Mod. Phys. Lett. {\bf A 6} (1991) 3385.
%
\bibitem{AGJ92}
  D. Atkinson, H.J. de Groot and P.W. Johnson,
  Intern. J. Mod. Phys. {\bf A 7} (1992) 7629.
%
\bibitem{Rakow91}
  P.E.L. Rakow,
Nucl. Phys. {\bf B 356} (1991) 27.
%
\bibitem{RS78}
  M. Reed and B. Simon,
  {\it Method of Modern Mathematical Physics, Vol. IV,
  Analysis of Operators} (Academic Press, New York, 1978).
%
\end{thebibliography}
\end{document}